\documentclass{appolb}
\usepackage{epsfig}
\usepackage{rotating}
\usepackage{epsfig}
\usepackage{amssymb}
\usepackage{amsmath}
\usepackage{amsfonts}
\usepackage{amssymb}
\usepackage{amscd}
\usepackage{amsthm}
\usepackage{graphicx}

\begin{document}
\title{
Jet and inclusive particle production in photon induced collisions
\thanks{Presented at PHOTON2005 Symposium, Warsaw, September 1-4, 2005}%
}
\author{Ji\v{r}\'{\i} Ch\'{y}la
\address{Institute of Physics, Academy of Sciences of the Czech Republic, Prague}
}
\maketitle

\vspace*{-0.8cm}
\begin{abstract}

\noindent
Recent progress in application of higher order QCD calculations to jet
and inclusive particle production in photon induced collisions is reviewed.
Attention
is paid to theoretical uncertainties of such calculations, particularly
those coming from the choice of renormalization and factorization
scales.

\vspace*{-0.3cm}
\end{abstract}

\vspace*{-0.3cm}
\section{Introduction}
\noindent
From the variety of hard processes involving initial state photon
I will discuss jet and inclusive particle production
in $\gamma$p and $\gamma\gamma$ collisions with emphasis on recent
phenomenological applications. I will concentrate on the discussion of
theoretical uncertainties because they are often bigger than experimental
errors thereby complicating the interpretation of data.

Let me start with brief recollection of basic facts concerning the structure
and hard collisions of protons and photons. For details see \cite{Maria}.
Parton distribution functions (PDF) of the photon depend on the
factorization scale, denoted $M$ below, and satisfy the system of
evolution equations
\begin{eqnarray}
\frac{{\mathrm d}\Sigma(x,M)}{{\mathrm d}\ln M^2}& =&
\delta_{\Sigma}k_q+P_{qq}\otimes \Sigma+ P_{qG}\otimes G,
\label{Sigmaevolution}
\\ \frac{{\mathrm d}G(x,M)}{{\mathrm d}\ln M^2} & =& k_G+
P_{Gq}\otimes \Sigma+ P_{GG}\otimes G, \label{Gevolution} \\
\frac{{\mathrm d}q_{\mathrm {NS}}(x,M)}{{\mathrm d}\ln M^2}& =&
\delta_{\mathrm {NS}} k_q+P_{\mathrm {NS}}\otimes q_{\mathrm{NS}},
\label{NSevolution}
\end{eqnarray}
where $\Sigma(x,M)$, $q_{\mathrm{NS}}$ and $G(x,M)$ stand for standard
singlet, nonsinglet and gluon distribution functions and the splitting
functions $P_{ij}$ and $k_i$ are given as power expansions in $\alpha_s(M)$.
For the proton the inhomogeneous splitting functions $k_q$ and $k_G$ are
absent.

General solution of the evolution equations
(\ref{Sigmaevolution}-\ref{NSevolution}) can be written as the sum
of a particular solution of the full inhomogeneous equations and the
general solution of the corresponding homogeneous ones. For instance,
the solution of (\ref{NSevolution}) including only
the lowest order terms $k_q^{(0)}$ and $P_{qq}^{(0)}$ and vanishing
at some factorization scale $M_0$, is given in momentum space as
\begin{equation}
q_{\mathrm {NS}}^{\mathrm {PL}}(n,M_0,M)=\frac{4\pi}{\alpha_s(M)}
\left[1-\left(\frac{\alpha_s(M)}{\alpha_s(M_0)}\right)^
{1-2P^{(0)}_{qq}(n)/\beta_0}\right]a_{\mathrm {NS}}(n),
\label{gp}
\end{equation}
\begin{equation}
a_{\mathrm {NS}}(n)\equiv \frac{\alpha}{2\pi\beta_0}
\frac{k_{\mathrm {NS}}^{(0)}(n)}{1-2P^{(0)}_{qq}(n)/\beta_0}.
\label{ans}
\end{equation}
It is often claimed that because of the presence of $\alpha_s$ in the
denominator of (\ref{gp}) quark distribution functions of the photon behave
as $\alpha/\alpha_s$. As argued in \cite{ja} this is misleading and PDF of
the photon actually start at the order $\alpha$. The standard assignment of
the order of PDF of the photon has important implications for the
definition of finite order approximations, but in this talk I will not
pursue this point further.

The cross section of inclusive hadron $h$ production in
$\gamma$p or $\gamma\gamma$ collisions has the generic form
\begin{equation}
\sigma(\gamma B\rightarrow hX)\propto \sum_{a,b,c}
D_{a/\gamma}(M_\gamma)\otimes D_{b/B}(M_B)\otimes
\sigma(ab\rightarrow cX)\otimes D_c^h(M_h)
\label{form}
\end{equation}
where the sum runs over quarks, gluons and photons (in the direct photon
channels) in the photon or proton. The hard partonic cross sections can be
expanded in powers of $\alpha_s$ as follows
\begin{equation}
\sigma(ab\rightarrow cX)=\alpha_s^k(\mu)\sigma^{LO}(ab\rightarrow cX)
+\alpha_s^{k+1}(\mu)\sigma^{NLO}(ab\rightarrow cX)+\cdots
\label{partonicXS}
\end{equation}
with $k=1$ for direct photon processes and $k=2$
for resolved ones. For jet production fragmentation
functions are replaced by some jet algorithm.

\section{Remarks on phenomenology}
\noindent
In (\ref{form}-\ref{partonicXS}) I have distinguished the
factorization scales of the beam particles from the fragmentation scale $M_h$
and the renormalization scale $\mu$. The latter appears only in perturbative
expansion of the partonic hard scattering cross sections (\ref{partonicXS}).
There is \underline{no theoretical reason} to identify any two or more of
these four scales as they come from the treatment of different singularities.
Unfortunately, in most phenomenological analyses all the four scales are
set equal and identified with some ``natural physical scale'' Q:
$\mu=M_\gamma=M_p=M_h=M=Q$.
However, choosing the renormalization scale does not fix $\alpha_s$
because $\alpha_s(\mu)$ depends beside the renormalization scale also on the
renormalization scheme (RS). Consequently, the same choice of the
renormalization scale gives different results in different RS! The choice
of the RS is in fact as important as that of renormalization scale $\mu$,
but there is no ``natural'' RS! The conventional procedure of
working in $\overline{\mathrm{MS}}$ RS is entirely ad hoc.
Choosing the scales and schemes should be based on the investigation of
the dependence of finite order approximants on renormalization and
factorization scales as independent parameters and should reflect the
possible existence of regions of local stability.

In comparing data with theoretical calculations the latter are quoted
with estimates of their ``theoretical uncertainty''. This quantity
usually
comprises several components: dependence on PDF and FF, hadronization
effects and scale and scheme variation. Hadronization corrections are
usually claimed to be small, but as we shall see in the case of jet and
inclusive particle production in $\gamma\gamma$ collisions it does not
have to be always the case and further investigation of this point is
certainly worth the efforts.

Whereas the uncertainty reflecting the dependence on PDF, FF and
hadronization is reasonable well-defined, the conventional way of
estimating the uncertainty due to the freedom of choice of scales and
scheme is definitely not. Identifying all scales with some ``natural
scale'' Q and varying this common scale $M$ typically in the interval
$Q/2\le M\le 2Q$ makes little sense. First, the results still depend
on the selected renormalization scheme, and, second, the chosen range
of multiplicative factor $(1/2,2)$ is again entirely arbitrary.

Another important aspect of the comparisons of theory with data concerns
the relation between LO MC event generators, like HERWIG, RAPGAP and
PYTHIA, and
NLO parton level calculations. Some of the features of full NLO QCD effects
are mimicked within LO MC event generators by means of parton showers and
noncollinear kinematics of initial state parton emissions. Moreover, LO MC
use different input (PDF, FF and $\alpha_s$) which were extracted in LO
global analysis from data, and so have a chance to describe also other data.

\section{Jet production in $\gamma$p and $\gamma^*$p collisions}
\noindent
Theory and phenomenology of jet production in (quasi)real
$\gamma$p collisions has recently been reviewed in \cite{Klasen}.
Extensive application to HERA data has lead to good agreement,
except for some IR sensitive quantities but suffers from non-negligible
scale dependence of existing QCD calculations.

In this talk I will concentrate on the phenomenological
application of QCD calculations to HERA data in the kinematic region of
moderate $Q^2$, where the transition region between photoproduction and
deep inelastic scattering takes place. This region is of particular interest
because of
the expected manifestation of effects due to BFKL dynamics.
Unfortunately, it is also the region, where playing with scales
does wonders and may easily mask the presence of new phenomena.
Although in this region the concept in resolved (virtual) photon does not
have to be introduced it turns out to be very useful phenomenologically.

From the NLO parton level Monte Carlo codes appropriate for this kinematic
region three are most frequently used in phenomenological
applications: DISENT \cite{DISENT}, NLOJET \cite{NLOJET} and JETVIP
\cite{JETVIP}. From them only JETVIP includes the contribution of resolved
virtual photon and only NLOJET offers the option of calculating three jet
production at the NLO.
\begin{figure}\unitlength=1mm
\begin{picture}(170,40)
\put(0,0){\epsfig{file=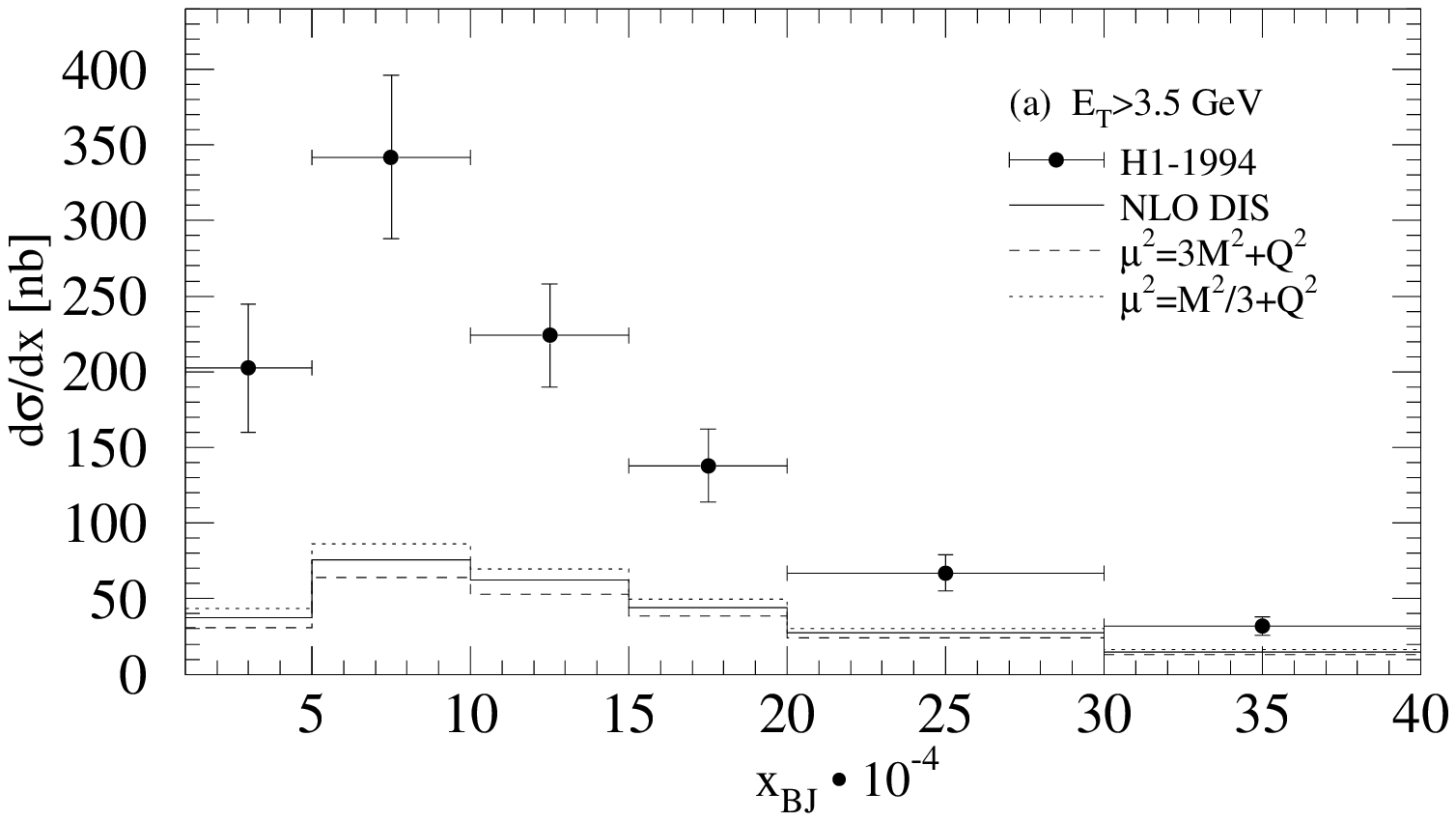,height=4.2cm}}
\put(63,0){\epsfig{file=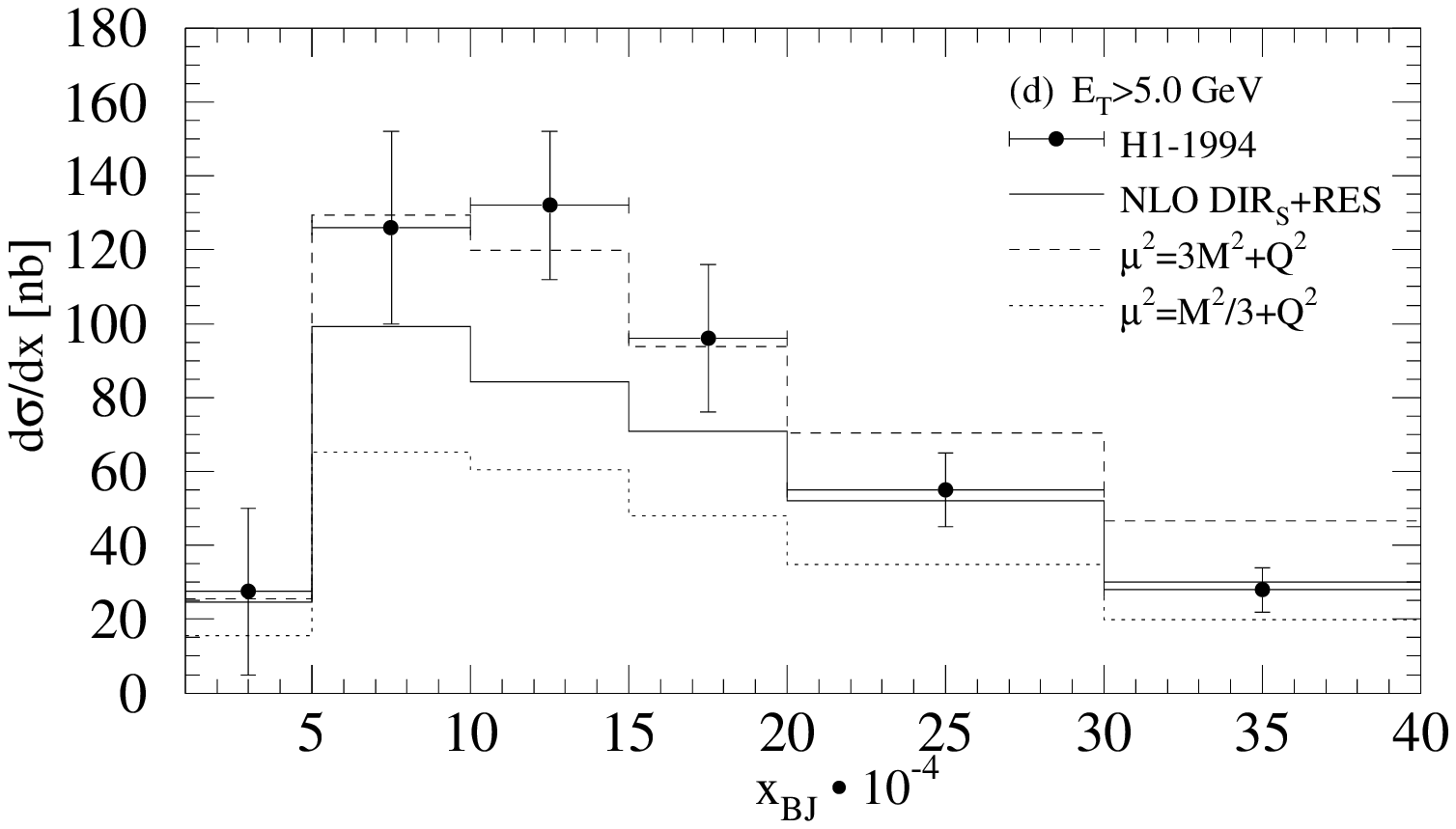,height=4.2cm}}
\end{picture}
\caption{Comparison of H1 data \cite{H1forward}
on forward jet production   with NLO QCD calculation
of \cite{kp}.}
\label{Forwardjets}
\end{figure}

Out of many interesting analyses of jet production I will briefly mention
two. The series of investigations of dijet production in the region
$Q^2\ll E_T^2$ has shown convincingly the importance of the contributions
of resolved virtual photon for the description of HERA data \cite{dijets}.
The relevance of the concept of resolved virtual photon as an approximate
method for including higher order direct photon contributions has
been studied in \cite{my} using NLOJET in the three jet mode.
These studies have also demonstrated
the importance of the contributions of the longitudinal
virtual photon.

Much attention has recently been paid
to forward jet production at HERA in the kinematic region where $Q^2/E_T^2$
is around unity. This region has been suggested \cite{Mueller} as suitable
place to look for manifestations of BFKL dynamics. However,
identification of these effects has turned out to be complicated by the
contributions of the resolved virtual photon, as well as by large scale
uncertainties of QCD calculations.

The ZEUS \cite{ZEUSforward} and H1 data \cite{H1forward} of forward jet
production in the kinematic region $0.5\le E^2_T/Q^2\le 2$, $E_T\ge 5$ GeV
and $p_z/E_p\ge 0,05$ have been analysed in NLO QCD using JETVIP
and including the resolved virtual photon contribution \cite{kp}. The
common scale $M$ was set to $M=\sqrt{\scriptstyle Q^2+E_T^2}$
and allowed to vary between $M=\sqrt{\scriptstyle Q^2+E_T^2/3}$  and
$M=\sqrt{\scriptstyle Q^2+3E_T^2}$.

The results, presented in Fig. \ref{Forwardjets} show that direct photon
contribution alone undershoots data significantly, but the inclusion of
NLO resolved photon contribution leads to nice agreement! However, as shown
by the dotted and dashed curves, this agreement relies crucially on the
chosen scale.

\section{Inclusive particle production in $\gamma^*$p collisions}
\noindent
Similarly as for jets the interest in this process is motivated
primarily by search for BFKL effects.
In \cite{Daleo} the recent H1 data on large $p_T$ forward
$\pi^0$ production \cite{H1pi0} in the region of moderate $Q^2$
were compared to QCD calculations including only the direct photon contribution.
Setting the common scale $\mu=M=\sqrt{\scriptstyle(Q^2+E_T^2)/2}$
\begin{figure}\unitlength=1mm
\begin{picture}(170,65)
\put(-5,0){\epsfig{file=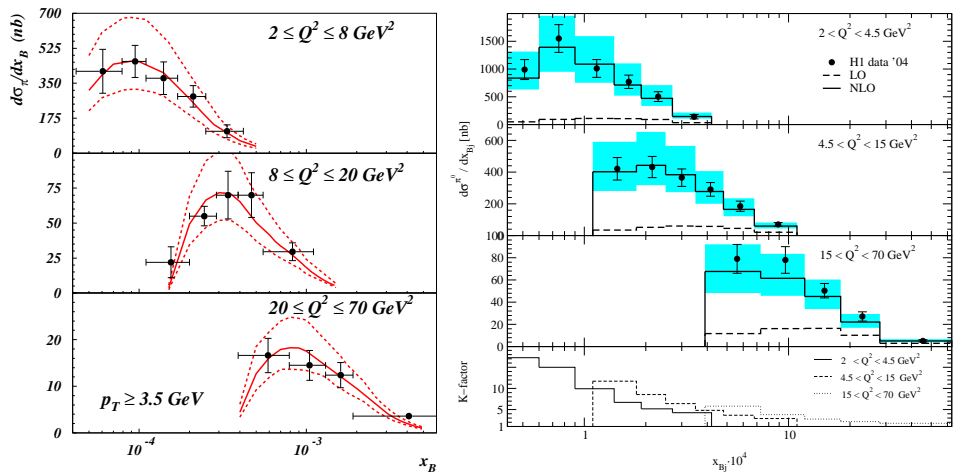,height=6.7cm}}
\put(55,0){
\begin{sideways}
\begin{sideways}
\begin{sideways}
{\epsfig{file=fig2b.eps,height=6.7cm}}
\end{sideways}
\end{sideways}
\end{sideways}}
\end{picture}
\caption{The H1 data on forward $\pi^0$ production \cite{H1pi0}
compared to calculations of \cite{Daleo} (left) and \cite{Kniehl}
(right), which include only direct photon contributions. The band
between two dashed curves in the left part and the shaded bands
at right correspond to the variation of the common scale in the
interval $\sqrt{\scriptstyle(Q^2+E_T^2)/4}\le
M\le \sqrt{\scriptstyle Q^2+E_T^2}$.}
\label{fpi0}
\end{figure}
the authors found a nice description of the H1 data for broad range of
values of $Q^2$ (see left part of Fig. \ref{fpi0}). However, they
noted strong dependence of their results on $M$. Varying their preferred
value of $M$ by factor of mere $\sqrt{2}$ their results change by a factor
of more than 2! This significant scale dependence in their
view ``suggests the presence of non-negligible NNLO effects'' and
together with large $K$-factors
`` restraining, for the moment, any empirical suggestion of dynamics
different to plain DGLAP evolution.''. I share their reservations.

The same H1 data on forward $\pi^0$ have been analyzed also in
\cite{Kniehl} using the same input and identical choice of scales as in
\cite{Daleo}. Not surprisingly they have also found a good agreement with
H1 data (see right plot of Fig. \ref{fpi0}) and factor of 2 difference
between their results corresponding to
$M\equiv\sqrt{\scriptstyle(Q^2+E_T^2)/4}$ and
$M\equiv\sqrt{\scriptstyle Q^2+E_T^2}$.
Compared to \cite{Daleo} the authors of \cite{Kniehl}
are more optimistic  and
claim that their ``default predictions, endowed with theoretical
uncertainties
estimated by moderate unphysical scale variations led to a satisfactory
description of the HERA data in the preponderant part of the accessed
phase space.'' In view of the arbitrariness of the standard choice of the
common scale I consider their conclusion difficult to justify.
\begin{figure}\unitlength=1mm
\begin{picture}(170,60)
\put(-2,0){\epsfig{file=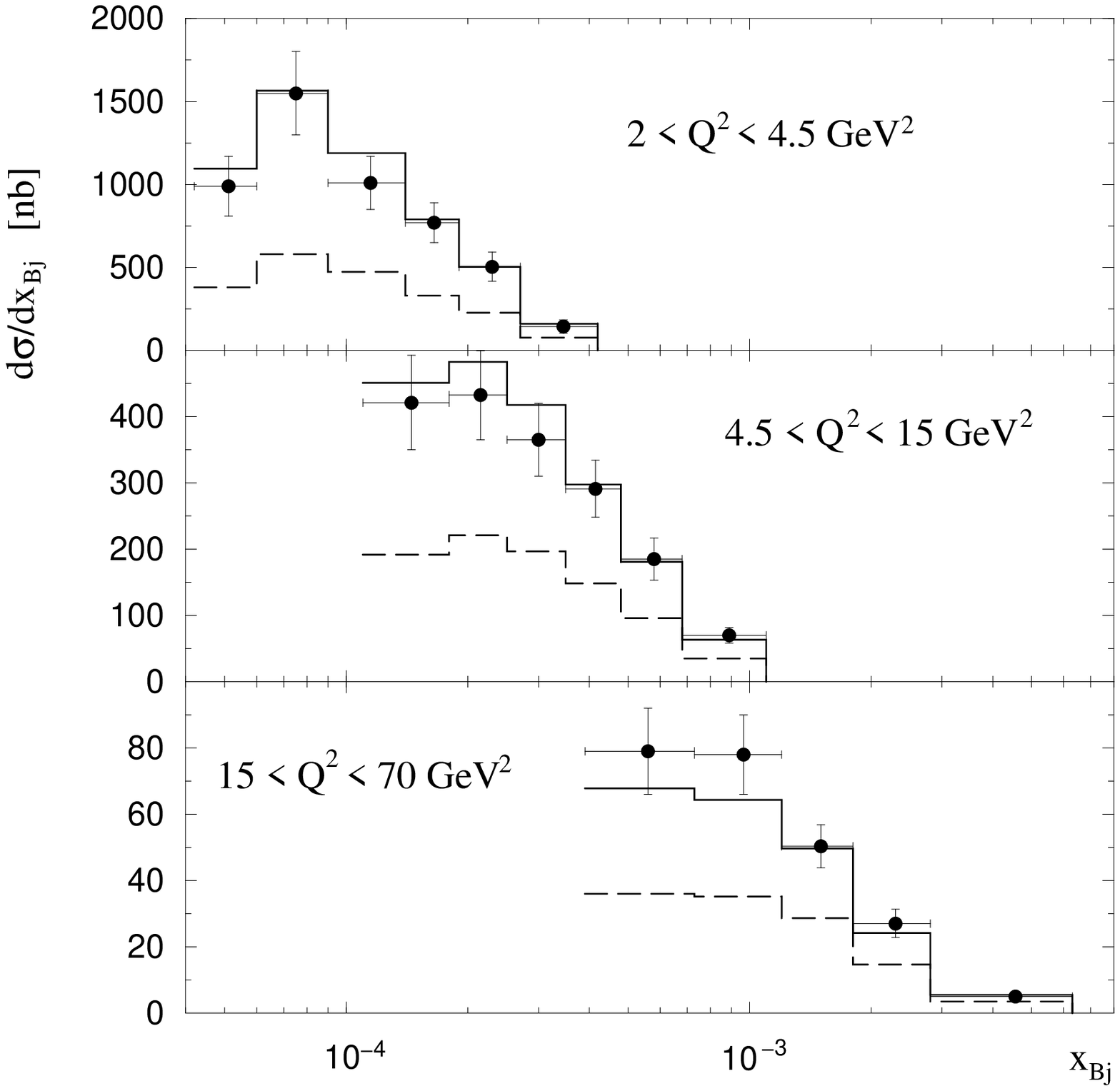,height=5.6cm}}
\put(63,0){\epsfig{file=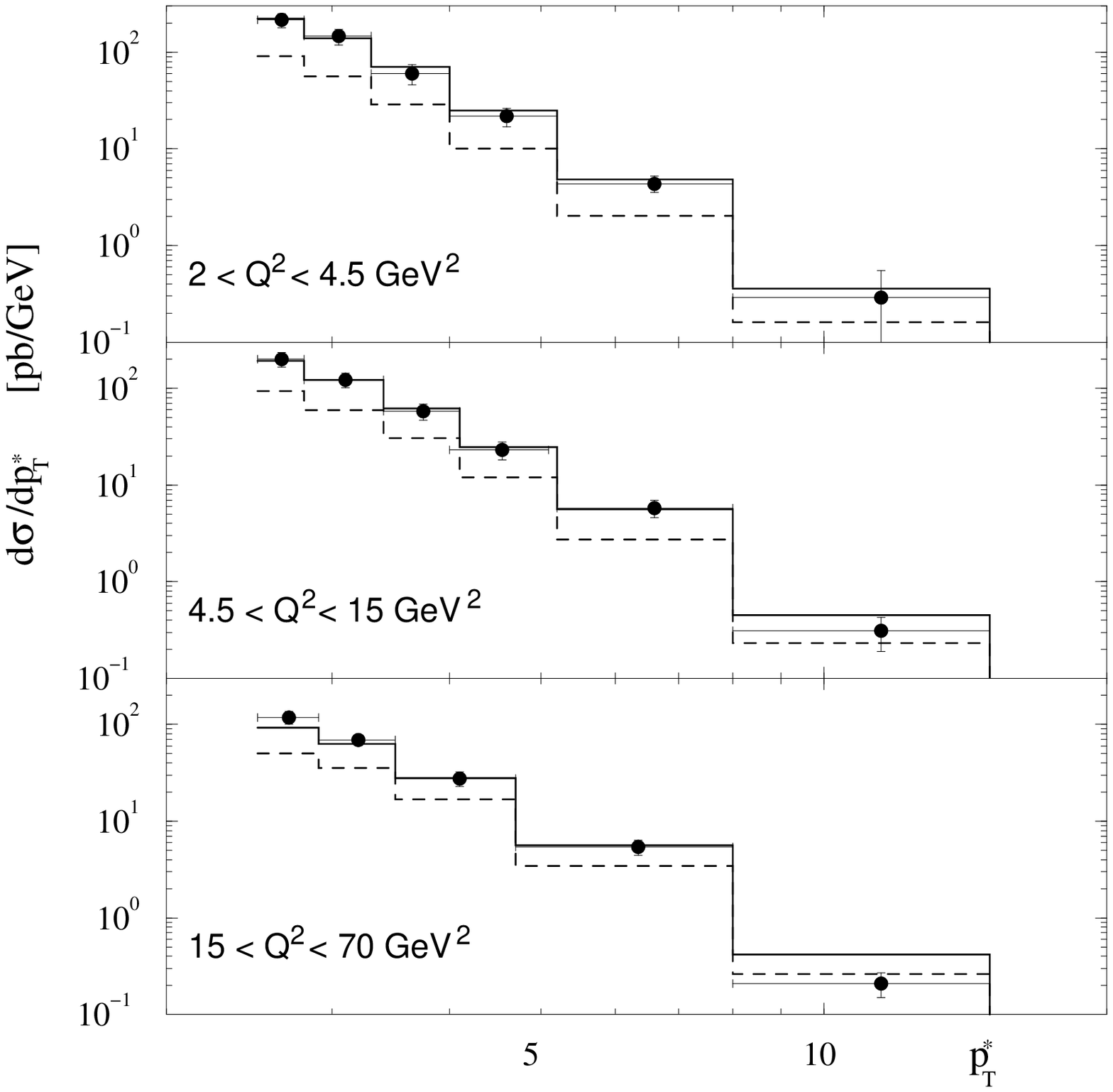,height=5.6cm}}
\end{picture}
\caption{The H1 data on forward $\pi^0$ as a function of
$x_{Bj}$ for $p_T> 2.5$ GeV (left) and of $p_T$ for three different
intervals in $Q^2$ compared to calculations of \cite{Michel2}
with (solid curves) and without (dashed curves) the resolved photon
contribution.}
\label{fpi0m}
\end{figure}

This is illustrated by the most recent analysis of the same H1
data performed in \cite{Michel2} taking into account
the contribution of the resolved virtual photon. This paper
includes the most detailed analysis of the
dependence on different scales, finding very different dependence on the
renormalization, factorization and fragmentation scales. In particular
``large instability is observed when varying independently the
renormalization and fragmentation scales.'' Default calculations (see Fig.
\ref{fpi0m}) have been performed for the common scale set to
$M\equiv\sqrt{\scriptstyle Q^2+E_T^2}$. Compared to \cite{Daleo} and
\cite{Kniehl} their common scale is bigger by a factor
$\sqrt{\scriptstyle 2}$ and thus their NLO direct contribution
substantially smaller with the resolved photon contribution filling the
gap! However, again large scale sensitivity
``prevents a really quantitative prediction for the single
pion inclusive distribution in the forward region.''

\section{Jets and inclusive particle production in $\gamma\gamma$
collisions}
\begin{figure}\unitlength=1mm
\begin{picture}(170,50)
\put(0,0){\epsfig{file=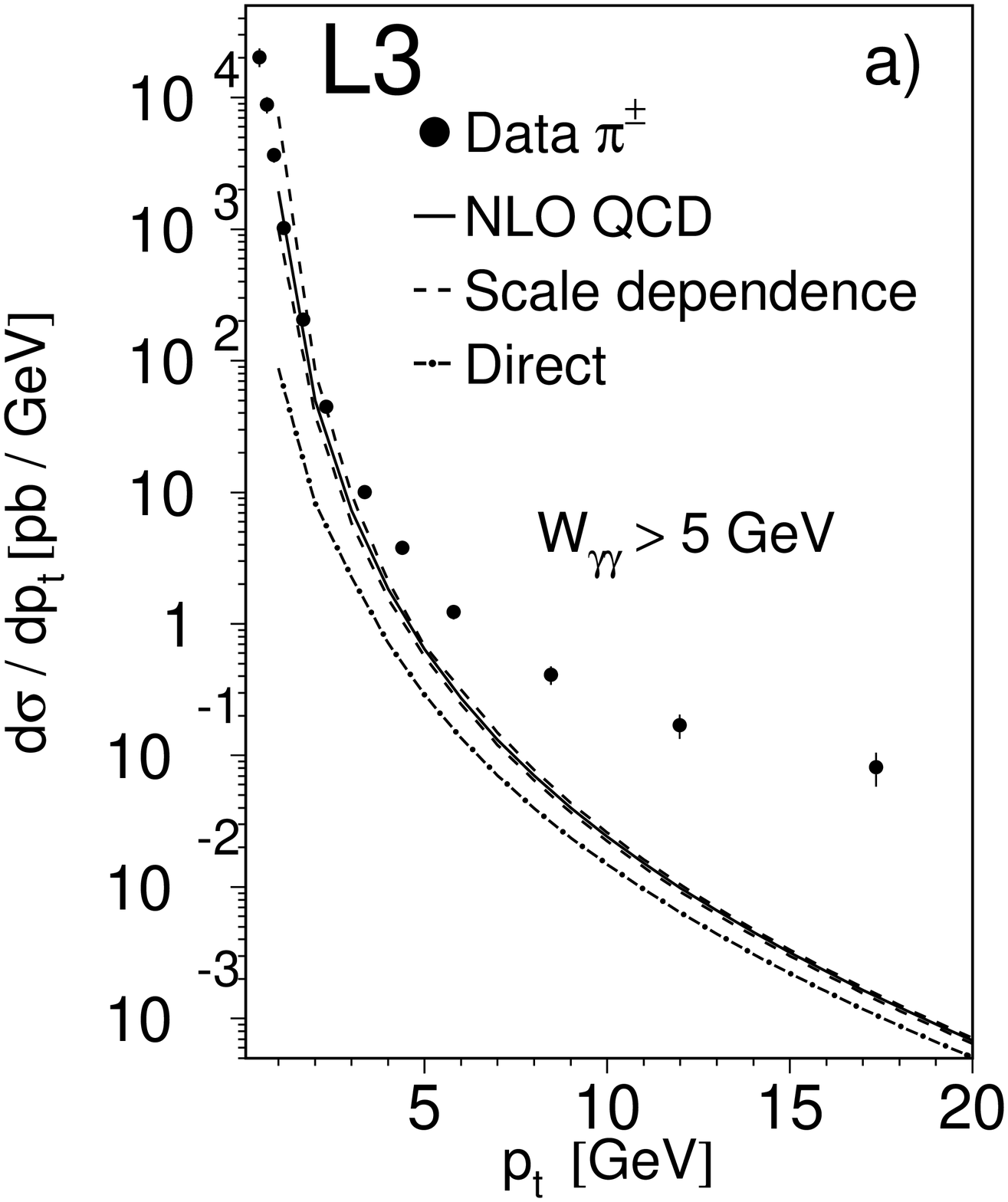,height=5cm}}
\put(42,0){\epsfig{file=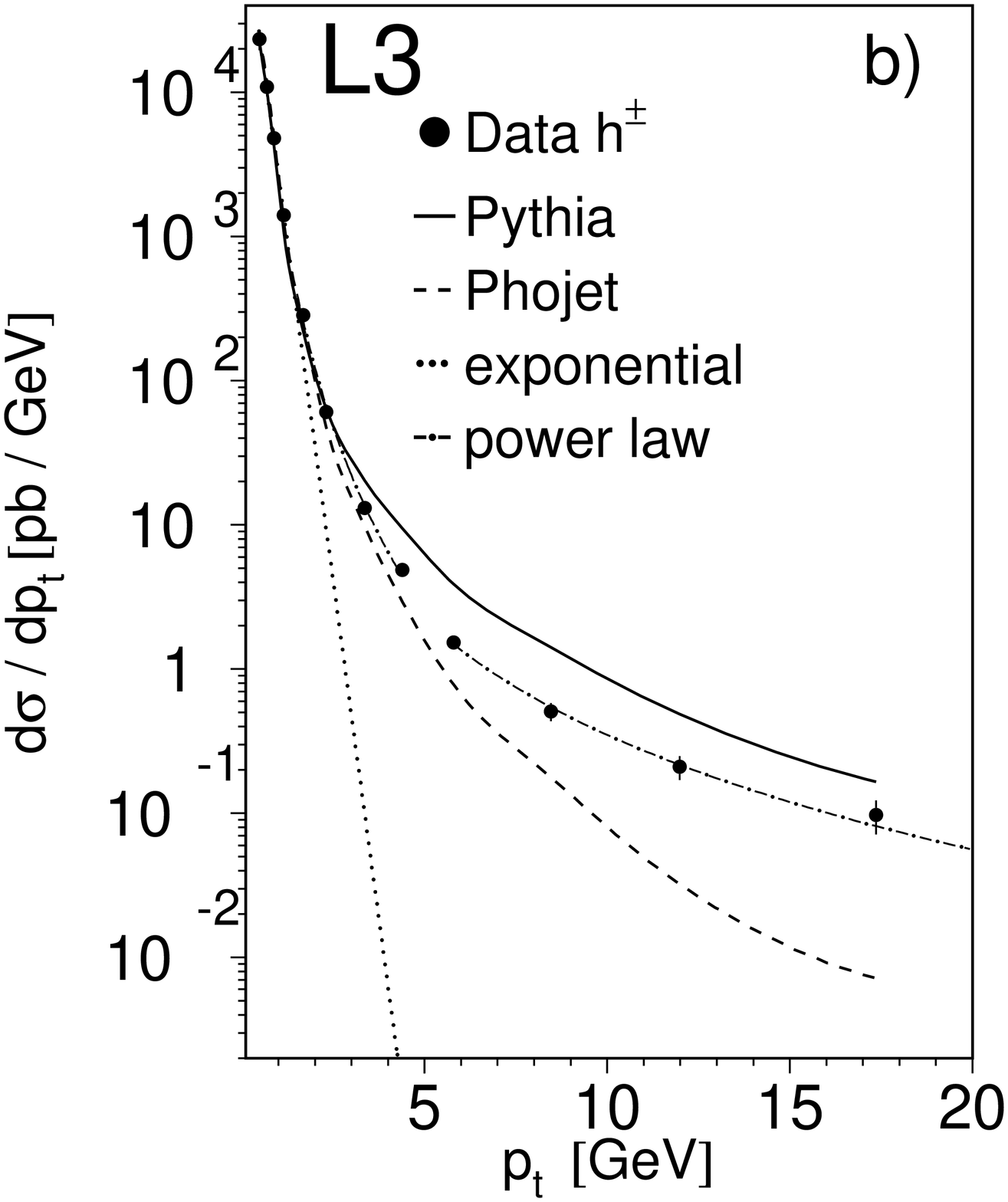,height=5cm}}
\put(84,0){\epsfig{file=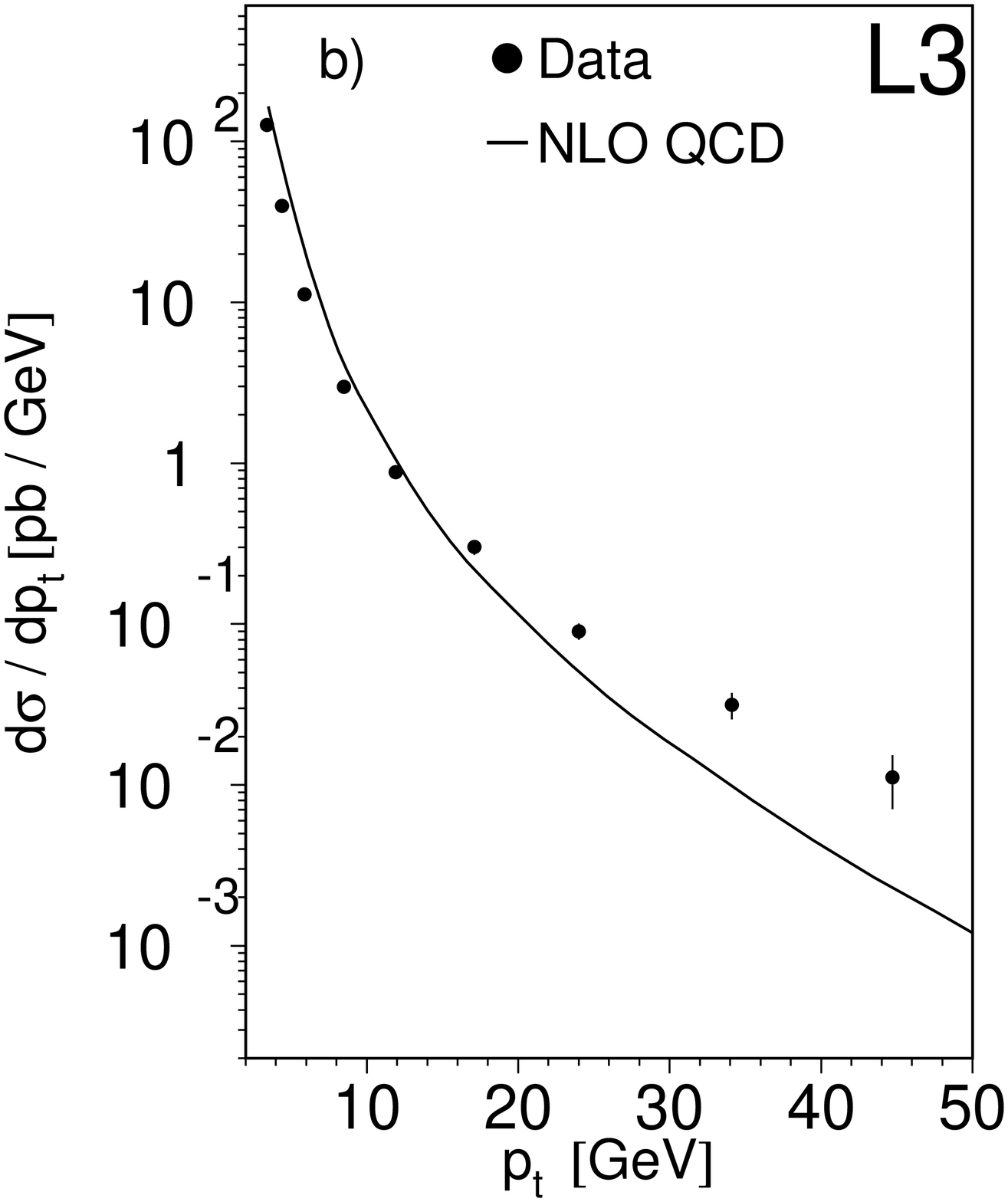,height=5cm}}
\end{picture}
\caption{Left: the comparison of L3 data on inclusive charged particle
production in $\gamma\gamma$ collisions \cite{l3incl} with the NLO parton
level calculation of \cite{Kniehl2}. Middle: the same data compared to
the predictions of PYTHIA and PHOJET LO event generators. Right: L3 data
on jet production compared to NLO QCD calculations of \cite{frix}.}
\label{gg}
\end{figure}

\noindent
Inclusive particle production in $\gamma\gamma$
collisions has been a challenge for perturbative QCD already at
PETRA and has become so even more at LEP2.
The recent L3 data on charged particle production in $\gamma\gamma$
collisions \cite{l3incl}, shown in Fig. \ref{gg}, are far above the NLO QCD
predictions \cite{Kniehl2} in the range of transverse momenta, where
the direct-direct contribution dominates and where there is thus
little chance that higher order QCD corrections could make any
difference.
There is, however, one aspect of the L3 analysis which deserves closer
attention. As shown in the middle plot of
Fig. \ref{gg}, precisely in the region where data
exceed by more than order of magnitude NLO QCD prediction there is a huge
difference between two LO MC event generators, PYTHIA and PHOJET, the
former even exceeding the data! Both these MC use the same LO cross sections,
which dominate the large $p_T$ region, but nevertheless lead to vastly
different shape of the distribution. This represents a warning that the
effects beyond purely perturbative stage of the process and where the MC
do differ, may play a crucial role for the tail of distributions like that
in Fig. \ref{gg}.

The disagreement between L3 data and QCD calculations extends, as
shown in the right plot of Fig. \ref{gg}, also to jet production.
Note that for jets both the shape and magnitude of L3 jet data
\cite{l3jets}
and NLO QCD predictions of \cite{frix} are significantly different.
However, also in this case PYTHIA lies above and PHOJET below the L3
data. The puzzle is still there.

\section{Summary and Conclusions}
\noindent
Jet and inclusive hadron production in $\gamma$p, $\gamma^*$p and
$\gamma\gamma$ collisions has provided wealth of new data for the
comparison with higher order QCD calculations. Unfortunately, in the
most interesting processes, like forward jet
and hadron production at HERA, where signals of
new physics are expected, theoretical predictions are burdened with
large uncertainties stemming primarily from strong dependence on
renormalization, factorization and fragmentation scales. The standard
procedure of identifying all these scales and setting them
equal to some physical scale characteristic for the process lacks
theoretical justification and, moreover, is ambiguous as it implicitly
assumes working in ad hoc chosen $\overline{\mathrm{MS}}$ RS.
Systematic investigation of scale dependence
of QCD calculations should be performed with
the aim of formulating better justified procedure for choosing the
various scales.

In theoretically seemingly clean case of jet and inclusive hadron
production in $\gamma\gamma$ collisions perturbative
QCD faces serious challenge to explain recent L3 data.

\end{document}